\documentclass[fleqn,usenatbib,]{mnras}
\usepackage{newtxtext,newtxmath}
\usepackage{natbib}
\usepackage{amsfonts}
\usepackage{subcaption}
\usepackage{lscape}
\usepackage{booktabs} 
\usepackage{scalerel}
\usepackage{xcolor}  
\newcommand{\HII}{H\protect\scaleto{\textup{II}}{1.2ex}}   
\bibpunct{(}{)}{;}{a}{}{,}

\usepackage[T1]{fontenc}

\DeclareRobustCommand{\VAN}[3]{#2}
\let\VANthebibliography\thebibliography
\def\thebibliography{\DeclareRobustCommand{\VAN}[3]{##3}\VANthebibliography}

\usepackage{graphicx}	
\usepackage{amsmath}	
\usepackage[normalem]{ulem} 

\title[Dense Cores and Filaments in M16]{Dense cores and filaments in M16: Enhanced formation efficiency in the stellar feedback-driven shell}

\author[N. Pervaiz et al.]{
Nageen Pervaiz,$^{1,3}$
Guo-Yin Zhang,$^{1}$\thanks{E-mail: zgyin@nao.cas.cn}
Alexander Men'shchikov,$^{2}$\thanks{E-mail: alexander.menshchikov@cea.fr}
and Jin-Zeng Li$^{1}$\thanks{E-mail: ljz@nao.cas.cn}
\\
$^{1}$National Astronomical Observatories, Chinese Academy of Sciences,
Beijing 100101, China\\
$^{2}$Universit\'{e} Paris-Saclay, Universit\'{e} Paris Cit\'{e}, CEA, CNRS, AIM,
91191 Gif-sur-Yvette, France\\
$^{3}$University of Chinese Academy of Sciences, Beijing 100049, China
}

\date{Accepted XXX. Received YYY; in original form ZZZ}

\pubyear{\the\year{}}

\begin{document}
\label{firstpage}
\pagerange{\pageref{firstpage}--\pageref{lastpage}}
 \maketitle

\begin{abstract}
We present a comprehensive analysis of dense cores and filamentary structures in the M16 Eagle Nebula using high-resolution ($11.7^{\prime\prime}$) surface density and temperature maps derived from \textit{Herschel} observations. Using the \textit{hires} algorithm for map construction and the \textit{getsf} method for source and filament extraction, we identified 233 cores and 111 filaments in this massive star-forming region. The filaments exhibit a median width of 0.4\,pc -- and a median linear density of 61\,$M_\odot$\,pc$^{-1}$, with 76\% being supercritical for gravitational fragmentation. Our radial analysis of the $\sim$60\,pc diameter shell driven by the central NGC 6611 cluster reveals strong enhancements in structure formation: filament formation efficiency (FFE) is 2.3 times higher within the shell (peaking at 22\%), while core density shows a concurrent 1.5-fold enhancement. The moderate correlation between core density and FFE ($r=0.67$) indicates coupled formation processes. Theoretical analysis demonstrates that observed surface densities exceed the critical threshold for fragmentation by a factor of $\sim$8, with a fragmentation timescale ($\sim$1.5--2.0\,Myr) comparable to the shell's dynamical age ($\sim$1.0--1.3\,Myr), indicating we are observing fragmentation in progress. These results reveal a hierarchical fragmentation sequence -- shell compression → filament formation → core formation -- providing clear observational evidence for positive feedback where massive star formation triggers secondary structure formation in the surrounding molecular cloud.
\end{abstract}




\section{Introduction}
\label{sect:intro}

In the prevailing paradigm of star formation, filamentary structures within molecular clouds play a central role as the primary conduits for mass assembly and the birth sites of prestellar cores \citep{Andre+2010, Menshchikov+2010, Andre+2014, Zhang+2024}. Observations, particularly from the \textit{Herschel} Space Observatory, have established that dense cores are almost exclusively embedded within filaments, suggesting a hierarchical fragmentation sequence: clouds fragment into filaments, which subsequently fragment into cores \citep{Konyves+2015, Zhang+2020}. While this framework is well-established in nearby, quiescent low-mass star-forming regions \citep{Motte+1998, Alves+2001, Kirk+2005}, its application to more distant and dynamic massive star-forming complexes -- where stellar feedback from O- and B-type stars is a dominant energetic force -- remains less constrained \citep[e.g.,][]{Watkins+2019}. A critical unanswered question is how this filamentary network itself originates and evolves under the influence of powerful stellar winds and radiation \citep{Pineda+2023, Hacar+2025}.

Massive stars exert a profound feedback on their natal environments \citep{Garay+Lizano1999,Krumholz+2014}. This feedback is often dichotomized into disruptive effects that disperse gas and suppress star formation, and triggering mechanisms that compress surrounding material and induce new generations of stars \citep{Elmegreen+Lada1977, Zinnecker+Yorke2007}. A classic signature of positive feedback is the ``collect-and-collapse'' process, where an expanding shell of gas, swept up by an \HII{} region or stellar wind, becomes gravitationally unstable and fragments into new dense structures \citep{Whitworth+1994, Elmegreen+1994, Dale+2007}. While numerous shell-like structures hosting young stellar objects have been identified \citep{Deharveng+2010, Zavagno+2010}, conclusive observational evidence linking the enhanced formation of the very filaments and cores -- the immediate progenitors of stars -- directly to the dynamical compression of a shell has been challenging to obtain.

The Eagle Nebula (M16) presents an ideal laboratory to investigate this link. It harbors the young cluster NGC 6611, whose massive members power a prominent \HII{} region and are believed to have driven the formation of a large-scale shell structure ($\sim$60 pc diameter) within the surrounding molecular cloud \citep{Hester+1996, Pound1998, Hill+2012, Karim+2025}. Previous studies of M16 with \textit{Herschel}, as part of the HOBYS key program, have revealed its filamentary nature and radial temperature gradient \citep{Hill+2012}. However, a high-resolution, systematic census of its dense core and filament population, specifically in the context of quantifying formation efficiencies within the feedback-driven environment, has been lacking.

To analyze M16, we employ the multiscale source and filament extraction method \textit{getsf} \citep{Menshchikov2021b}, which separates the structural components of sources, filamentary structures, and background before applying a consistent extraction method. A key feature distinguishing \textit{getsf} from other extraction methods is that it has no free parameters. Its only user-defined parameter is the maximum spatial scale of the sources or filaments of interest, which is constrained by the observed images and sets an upper limit on the range of scales in the spatial decomposition.

This paper is organized as follows. Section~\ref{sect:obs} describes the \textit{Herschel} observations and data. Section~\ref{sect:analysis} presents our methodology for constructing high-resolution maps and extracting cores and filaments, along with the resulting catalogs. Section~\ref{sect:discussion} analyzes the physical properties of cores and filaments, examines the large-scale shell structure and its role in triggering enhanced structure formation, and discusses the hierarchical fragmentation sequence from shell to filaments to cores. Section~\ref{sect:conclusion} summarizes our main findings.

\section{Observations and data}
\label{sect:obs}

We use archival \textit{Herschel} Space Observatory observations of M16 obtained as part of the HOBYS (\textit{Herschel} imaging survey of OB Young Stellar objects) key program \citep{Motte+2010}. The observations were conducted in two epochs: 2010 March 24 (PACS only) and 2010 September 11--12 (PACS and SPIRE in parallel-scan mode).

The PACS instrument (Photodetector Array Camera and Spectrometer; \citealt{Poglitsch+2010}) provided imaging at 70 and 160 $\mu$m, while the SPIRE instrument (Spectral and Photometric Imaging Receiver; \citealt{Griffin+2010}) mapped at 250, 350, and 500 $\mu$m. The parallel-scan observations used a scan speed of 20$^{\prime\prime}$\,s$^{-1}$ with two orthogonal scan directions to minimize striping artifacts, covering approximately 1.5 square degrees.

We obtained fully calibrated Level 2.5 data products from the \textit{Herschel} Science Archive. The PACS maps were processed with Standard Product Generation (SPG) software version 14.2.0, and the SPIRE maps with SPG version 14.1.0, using their respective calibration trees. These products include instrumental response corrections, astrometric calibration, and final map projection.

\section{Analysis and results}
\label{sect:analysis}

\subsection{High-resolution surface density and temperature}
\label{sect:hires_maps}

To derive accurate surface densities from the \textit{Herschel} images, zero-level offsets must be determined for each of the five wavebands. We obtained these offsets by comparing the \textit{Herschel} images with simulated emission maps derived from \textit{Planck} data. Using the \textit{Planck} all-sky maps of dust optical depth at 353 GHz (850\,$\mu$m) and dust temperature at 5$^{\prime}$ resolution, we generated simulated intensity maps at the \textit{Herschel} wavelengths (70, 160, 250, 350, and 500 $\mu$m) assuming optically thin dust emission with a fixed opacity spectral index $\beta = 2$. Each \textit{Herschel} image was smoothed to the \textit{Planck} resolution and compared with the corresponding simulated map to determine the offset value \citep{Bernard+2010}.

We used the \textit{hires} algorithm \citep{Menshchikov2021b} to create high-resolution surface density and temperature maps from the \textit{Herschel} multiwavelength data. First, temperature and surface density maps are derived at three different resolutions: 36.3$^{\prime\prime}$ using images at 160, 250, 350, and 500 $\mu$m; 24.9$^{\prime\prime}$ using images at 160, 250, and 350 $\mu$m; and 18.2$^{\prime\prime}$ using images at 160 and 250 $\mu$m. These initial temperature maps are then used with the full set of observed images (70--500 $\mu$m) to compute surface densities at each wavelength. The final high-resolution surface density image is constructed by combining information from all resolutions \citep{Menshchikov2021b}:
\begin{equation}
\mathcal{D}_{{O}_{\mathrm{H}}}= \mathcal{D}_{O_{500}}+\sum^{500}_{\lambda=\lambda_{\mathrm{H}}}
\max\left(\delta\mathcal{D}_{O_{\lambda}2},\delta\mathcal{D}_{O_{\lambda}3},\delta\mathcal{D}_{O_{\lambda}4}\right),
\label{superdens}
\end{equation}
where $\mathcal{D}_{{O}_{\mathrm{H}}}$ is the high-resolution surface density at resolution ${O}_{\mathrm{H}}$, $\mathcal{D}_{O_{500}}$ is the surface density at 36.3$^{\prime\prime}$ resolution, and $\delta\mathcal{D}_{O_{\lambda}\{2|3|4\}}$ are differential terms containing higher-resolution information from shorter wavelengths.

The corresponding high-resolution temperature image is computed by numerically inverting the Planck function \citep{Menshchikov2021b}:
\begin{equation}
\mathcal{T}_{\!{O}_{\mathrm{H}}}= B^{-1}_{{\nu}_{\mathrm{H}}}
\left(\frac{\mathcal{I}_{{\nu}_{\mathrm{H}}}}{\mathcal{D}_{{O}_{\mathrm{H}}} \kappa_{{\nu}_{\mathrm{H}}} \eta \mu m_{\mathrm{H}}}\right),
\label{supertemp}
\end{equation}
where $\mathcal{I}_{{\nu}_{\mathrm{H}}}$ is the observed intensity at the highest resolution wavelength, $\kappa_{{\nu}_{\mathrm{H}}}$ is the dust opacity, $\eta=0.01$ is the dust-to-gas mass ratio, $\mu=2.8$ is the mean molecular weight per H$_2$ molecule, and $m_{\mathrm{H}}$ is the mass of the hydrogen atom.

\begin{figure*}
    \centering
    \includegraphics[width=\hsize]{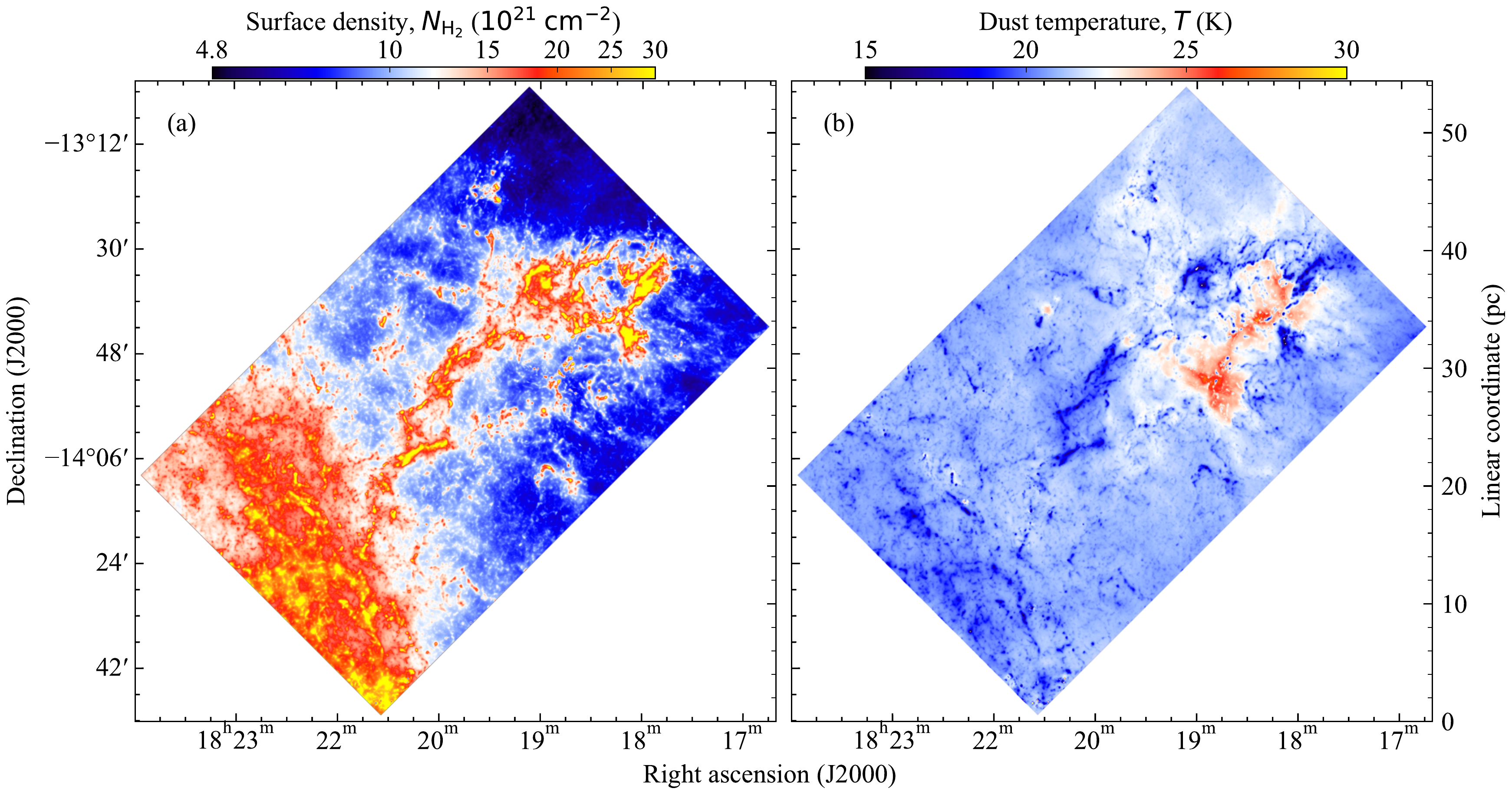}
     \caption{
     High-resolution ($11.7^{\prime\prime}$) maps of surface density (\textit{left}) and temperature (\textit{right}) of the M16 molecular cloud, derived from the \textit{Herschel} multiwavelength observations using the \textit{hires} algorithm. The surface density is shown in units of H$_2$ cm$^{-2}$ and the temperature in K.
     The right secondary axis on each panel shows the corresponding linear scale in parsecs, providing a physical size reference.
     }
    \label{fig:m16_maps}
\end{figure*}

In this study, we used the surface density and temperature images $\mathcal{D}_{11.7^{\prime\prime}}$ and $\mathcal{T}_{11.7^{\prime\prime}}$ at 11.7$^{\prime\prime}$ effective resolution, corresponding to 0.097 pc at the distance of M16 ($d\approx 1700$ pc, \citealt{Zucker+2020}; Fig.~\ref{fig:m16_maps}). These maps provide the observational basis for extracting and analyzing dense cores and filaments. Core and filament temperatures were obtained from the temperature map as
\begin{equation}
T_{\text{c}} = \mathcal{T}_{11.7^{\prime\prime}}(x_{\text{p}}, y_{\text{p}}), \,\,\, T_{\text{f}} = \operatorname{median}\left(\mathcal{T}_{11.7^{\prime\prime}}(x_{i}, y_{i})\right),
\label{eq:core_temp}
\end{equation}
where $(x_{\text{p}}, y_{\text{p}})$ are the coordinates of the core peak and $(x_i, y_i)$ are the coordinates of pixels along the filament skeleton.

\subsection{Source and filament extraction}

We used the multiscale, multiwavelength method \textit{getsf}\footnote{\url{http://irfu.cea.fr/Pisp/alexander.menshchikov/}} \citep{Menshchikov2021b} to extract sources and filaments. For detailed descriptions of the method and benchmarks demonstrating its performance, readers are referred to \citet{Menshchikov2021b,Menshchikov2021a}. Filaments were extracted from the 11.7$^{\prime\prime}$ surface density map, while sources were detected and measured in all \textit{Herschel} images (70--500 $\mu$m) as well as the surface density map.

The only free parameter of \textit{getsf} is the maximum size of the structures to be extracted, which is constrained by the footprint radius of the largest source or filament of interest in the image. This parameter sets a practical upper limit on the largest scales to be processed during the spatial decomposition, structural component separation, and detection of sources and filaments. The actual upper limit for the decomposition is set to four times this maximum size (Appendix~B in \citealp{Menshchikov2021b}). 
We estimated the largest sizes of sources and filaments by visually inspecting the images and adopted the maximum sizes of 25 and 30$^{\prime\prime}$, respectively.

To distinguish protostars from starless cores, we identified protostars using the 70 $\mu$m \textit{Herschel} images. Due to variations in angular resolution and dust temperature, the observed peak emission of relatively hot protostars and their parent cores (with colder dust) can appear at slightly different locations across different wavelengths. Protostars are more prominent at shorter wavelengths such as 70 $\mu$m due to internal heating from accretion energy \citep{Andre+2010, Hennemann+2010, Konyves+2015}.

The total mass of the M16 cloud in the \textit{Herschel}-observed region (excluding the Galactic plane region, which does not belong to M16) is $M_{\text{tot}} = 2.5 \times 10^{5} \, M_{\odot}$, obtained by integrating the entire surface density map (Fig.~\ref{fig:m16_maps}). The \textit{getsf} method separates the structural components into sources, filaments, and background. We find that the extracted sources contain 1\% of $M_{\text{tot}}$, the filaments contain 11\%, and the background contains 88\%.

\subsection{Selection and classification of cores}
\label{sec:core_selection_classification}

To ensure reliable physical measurements of each core, we applied the following selection criteria recommended by \citet{Menshchikov2021b} to the high-resolution surface density map $\mathcal{D}_{11.7^{\prime\prime}}$:
\begin{eqnarray}
\left.\begin{aligned}
&{\Xi > 1} \,\land\, {\Gamma > 1} \,\land\, {\Omega > 2} \,\land\, 
{\Psi > 2} \,\land\, \\
&{A < 2 B} \,\land\, {A_{{\rm F}} > 1.15 A},
\end{aligned}\right.
\label{acceptable} 
\end{eqnarray}
where $\Xi$ is the detection significance, $\Gamma$ is the monochromatic goodness, $\Omega = F_{\mathrm{P}}/F_{\mathrm{P,err}}$ and $\Psi = F_{\mathrm{T}}/F_{\mathrm{T,err}}$ are the peak and integrated surface density signal-to-noise ratios, $A$ and $B$ are the major and minor half-maximum sizes, and $A_{{\rm F}}$ is the major axis of the footprint. 
The monochromatic goodness $\Gamma$ quantifies the quality of a source detection in a single band, here the 11.7$^{\prime\prime}$ surface density map. It combines the detection significance $\Xi$, the peak and integrated signal-to-noise ratios ($\Omega$ and $\Psi$), and the source elongation ($B/A$) as $\Gamma = (\Xi/5) \times (\sqrt{\Omega\Psi}/2) \times (B/A)$, normalized such that $\Gamma > 1$ indicates an acceptable source extraction \citep[Eqs.~(40)--(42) in][]{Menshchikov2021b}.

Additionally, we required all core parameters to be positive and excluded cores in the Galactic plane region. After applying these selection criteria, 233 reliable cores remained for further analysis. We classified these cores into three categories based on their evolutionary state and gravitational boundedness. Protostellar cores were identified by cross-matching with an independent protostellar catalog derived from 70\,$\mu$m emission \citep[following the method of][]{Konyves+2015}, yielding 75 protostellar cores. The remaining 158 cores are starless.

The mass of each core was determined directly from the high-resolution surface density by summing over the core footprint:
\begin{equation}
M_{\text{c}} = \mu m_{\mathrm{H}} \Delta^2 \sum_{i} \mathcal{D}_{11.7^{\prime\prime}, i},
\label{eq:core_mass}
\end{equation}
where $\Delta$ is the pixel size (3$^{\prime\prime}$), $\mu=2.8$ is the mean molecular weight per H$_2$ molecule, $m_{\mathrm{H}}$ is the mass of the hydrogen atom, and the summation is over all pixels $i$ within the core footprint as determined by \textit{getsf}.

To assess the gravitational boundedness of the starless cores, we employed the critical mass $M_{\rm BE}$ of a Bonnor-Ebert (BE) sphere \citep{Ebert1955, Bonnor1956}:
\begin{equation}
M_{\rm BE} = 2.4 R_{\text{c}} c_{\rm s}^2 / G, \quad c_{\rm s} = \sqrt{k_{\rm B} T_{\text{c}} / (\mu m_{\text{H}})},
\end{equation}
where $R_{\text{c}} = 0.5\sqrt{A_{\rm F} B_{\rm F}}$ is the core boundary radius (from the major and minor axes of its elliptical footprint), $c_{\rm s}$ is the isothermal sound speed, $k_{\rm B}$ is the Boltzmann constant, and $G$ is the gravitational constant. The ratio $\alpha_{\rm BE} = M_{\rm BE} / M_{\text{c}}$ indicates gravitational boundedness: cores with $\alpha_{\rm BE} \leq 2$ are considered prestellar (gravitationally bound), while those with $\alpha_{\rm BE} > 2$ are unbound. Among the 158 starless cores, 115 are prestellar and 43 are unbound. 
We note that $\alpha_{\mathrm{BE}}$ is used solely as a threshold to separate gravitationally bound starless cores (prestellar) from unbound ones; protostellar cores are identified independently via 70\,$\mu$m emission, and their $\alpha_{\mathrm{BE}}$ values are reported only for reference.

\begin{figure*}
    \centering
    \includegraphics[width=0.95\hsize]{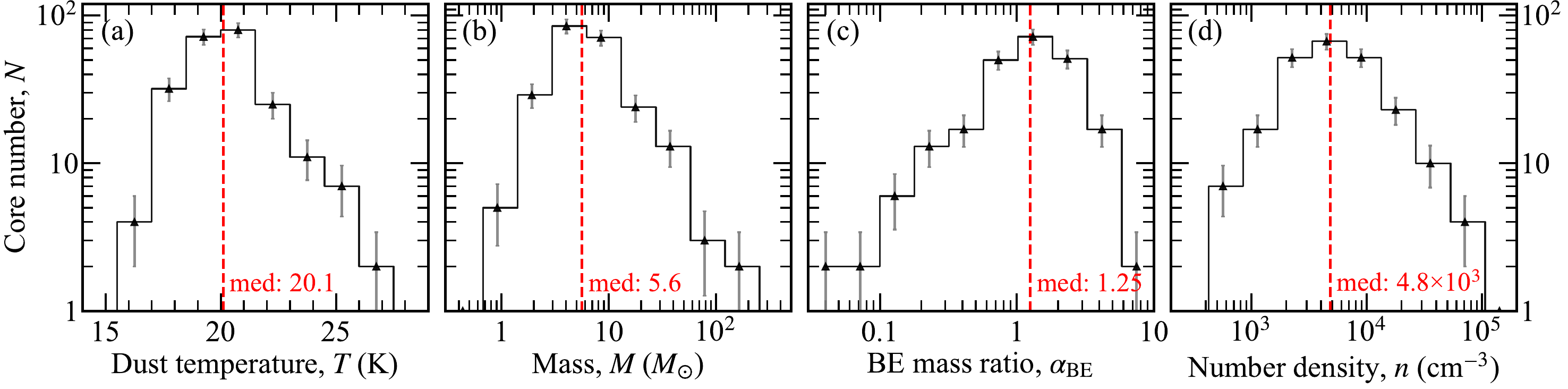}
     \caption{Distributions of physical properties for 233 dense cores in M16. Histograms show: (\textit{a}) core temperature $T_{\rm c}$, (\textit{b}) core mass $M_{\rm c}$, (\textit{c}) Bonnor-Ebert mass ratio $\alpha_{\text{BE}} = M_{\rm BE}/M_{\rm c}$, and (\textit{d}) volume-averaged H$_{2}$ number density $n_{\rm c}$. Red dashed vertical lines indicate median values. Cores with $\alpha_{\text{BE}} < 2$ are considered gravitationally bound.}
\label{fig:core_properties}
\end{figure*}

The physical properties of the 233 cores are summarized in Fig.~\ref{fig:core_properties} and Table~\ref{tab:core_statistics}. The population exhibits median values of: dust temperature $\tilde{T}_{\rm c}= 20.1$\,K, mass $\tilde{M}_{\rm c}= 5.6$\,$M_\odot$, Bonnor-Ebert mass ratio $\tilde{\alpha}_{\rm BE}= 1.25$, and volume number density $\tilde{n}_{\rm c}= 4.8\times 10^3$\,cm$^{-3}$. 

\begin{figure}
    \centering
    \includegraphics[width=0.95\hsize]{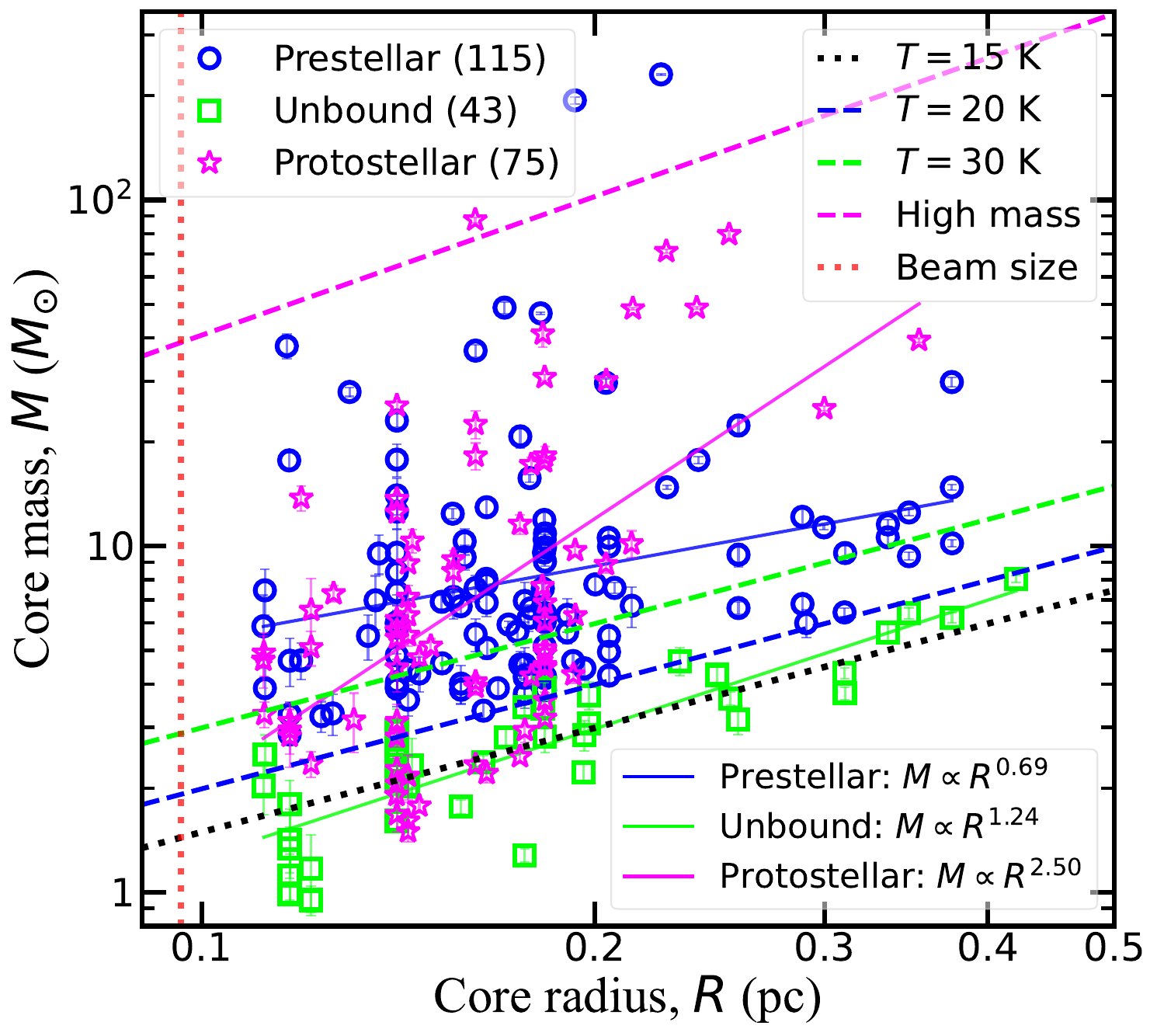}
    \caption{
    Mass-radius diagram for 233 dense cores in M16. Symbols indicate core classifications: unbound cores (green squares), prestellar cores (blue circles), and protostellar cores (red stars). Coloured solid lines show the power-law fits for each population. Dashed black, blue, and green curves show critical Bonnor-Ebert masses for temperatures of 15, 20, and 30 K, respectively. The magenta dashed line shows the empirical threshold for high-mass star formation from \citet{Kauffmann+2010}. The vertical red dotted line at 0.096 pc indicates the spatial resolution limit ($11.7^{\prime\prime}$ at 1700 pc distance).
    }
    \label{fig:mass_size}
\end{figure}

The mass-radius distribution (Fig.~\ref{fig:mass_size}) reveals systematic differences between core types. Prestellar cores are significantly more massive than unbound cores, with two positioned above the empirical threshold for massive star formation proposed by \citet{Kauffmann+2010}. A substantial fraction of cores -- including both prestellar and protostellar populations -- have masses exceeding 8\,$M_\odot$, the approximate threshold for forming B-type or more massive stars. The most massive core reaches 231\,$M_\odot$ (see Appendix~\ref{app:core_catalog}), confirming M16's status as an active massive star-forming region capable of producing high-mass stellar systems. To quantify the structural scaling of different core populations, we fitted a power-law relation $M_{\rm c} \propto R_{\rm c}^{\gamma}$ to the mass–radius data for each core type using linear regression in logarithmic space (Fig.~\ref{fig:mass_size}). The derived power-law indices $\gamma$ (and coefficients of determination $R^2$) are $0.69$ ($0.07$) for prestellar cores, $1.24$ ($0.63$) for unbound cores, and $2.50$ ($0.30$) for protostellar cores.

\begin{table*}
\centering
\caption{Statistical properties of dense cores in M16.}
\label{tab:core_statistics}
\begin{tabular}{lcccc}
\toprule
Parameter & All cores ($N=233$) & Prestellar ($N=115$) & Unbound ($N=43$) & Protostellar ($N=75$) \\
\midrule
$T_{\rm c}$ (K) & & & & \\
Mean $\pm$ stdev & $20.3 \pm 1.9$ & $19.8 \pm 1.6$ & $21.6 \pm 1.7$ & $20.3 \pm 2.0$ \\
Median & 20.1 & 19.7 & 21.1 & 20.0 \\
Range & 15.8--27.3 & 15.8--25.8 & 19.5--27.3 & 16.8--26.5 \\
\midrule
$M_{\rm c}$ ($M_\odot$) & & & & \\
Mean $\pm$ stdev & $11.0 \pm 22.3$ & $13.1 \pm 27.9$ & $2.9 \pm 1.5$ & $12.5 \pm 17.3$ \\
Median & 5.6 & 6.8 & 2.6 & 5.5 \\
Range & 0.5--230.6 & 2.9--230.6 & 0.5--8.1 & 1.5--87.7 \\
\midrule
$R_{\rm c}$ (pc) & & & & \\
Mean $\pm$ stdev & $0.18 \pm 0.06$ & $0.19 \pm 0.06$ & $0.19 \pm 0.08$ & $0.16 \pm 0.04$ \\
Median & 0.17 & 0.18 & 0.16 & 0.16 \\
Range & 0.11--0.42 & 0.11--0.38 & 0.11--0.42 & 0.11--0.35 \\
\midrule
$\alpha_{\text{BE}} = M_{\rm BE}/M_{\rm c}$ & & & & \\
Mean $\pm$ stdev & $1.53 \pm 1.28$ & $1.04 \pm 0.50$ & $3.21 \pm 1.70$ & $1.33 \pm 1.02$ \\
Median & 1.25 & 1.01 & 2.68 & 1.05 \\
Range & 0.04--11.98 & 0.04--1.97 & 2.00--11.98 & 0.06--4.16 \\
\midrule
$n_{\rm c}$ (cm$^{-3}$) & & & & \\
Mean $\pm$ stdev & $(9.0 \pm 14.0) \times 10^3$ & $(10.7 \pm 17.1) \times 10^{3}$ & $(2.5 \pm 1.5) \times 10^{3}$ & $(10.2 \pm 11.6) \times 10^{3}$ \\
Median & $4.8 \times 10^3$ & $5.7 \times 10^{3}$ & $2.3 \times 10^{3}$ & $6.6 \times 10^{3}$ \\
Range & $4.5\times10^{2}$--$1.1\times10^{5}$ & $8.0\times10^{2}$--$1.1\times10^{5}$ & $4.5\times10^{2}$--$7.4\times10^{3}$ & $1.9\times10^{3}$--$8.5\times10^{4}$ \\
\bottomrule
\end{tabular}
\end{table*}

\subsection{Selection and measurement of filaments}
\label{sect:filament_measurement}

Filaments are often blended with neighboring structures, embedded sources, and complex background emission, which presents difficulties -- particularly at limited angular resolutions \citep{Menshchikov2021b}. Selecting well-isolated filaments is essential for reliable parameter measurements. In our analysis, we focused on straight, non-branching filaments identified through \textit{getsf}-detected skeletons in the 11.7$^{\prime\prime}$ resolution surface density map, where sources and backgrounds have been subtracted.

Our measurements of filament properties followed the approach described by \citet{Menshchikov2021b}. We extracted radial profiles on each side of the filament crest, averaging the measurements when available on both sides or using one-sided measurements when only one side is suitable. We identified 111 well-defined filaments with lengths greater than 1 pc that are sufficiently isolated from neighboring structures to minimize blending effects.

For each filament, we measured several key physical parameters. The filament width $W$ is measured as the full width at half maximum (FWHM) from the radial profile. The crest surface density $N_{\mathrm{H}_2}$ is obtained as the median value along the skeleton. The linear mass density $\Lambda$ (mass per unit length) is calculated by integrating the surface density profile from the center to the edge of the filament:
\begin{equation}
\Lambda = 2 \mu m_{\rm H} \int_0^{R_{\text{f}}} N_{\mathrm{H}_2}(r) \, {\rm d}r,
\label{eq:linear_density}
\end{equation}
where $R_{\text{f}}$ is the filament footprint radius, $\mu = 2.8$ is the mean molecular weight per H$_2$ molecule, and $m_{\rm H}$ is the mass of the hydrogen atom.

\begin{figure*}
    \centering
    \includegraphics[width=\textwidth]{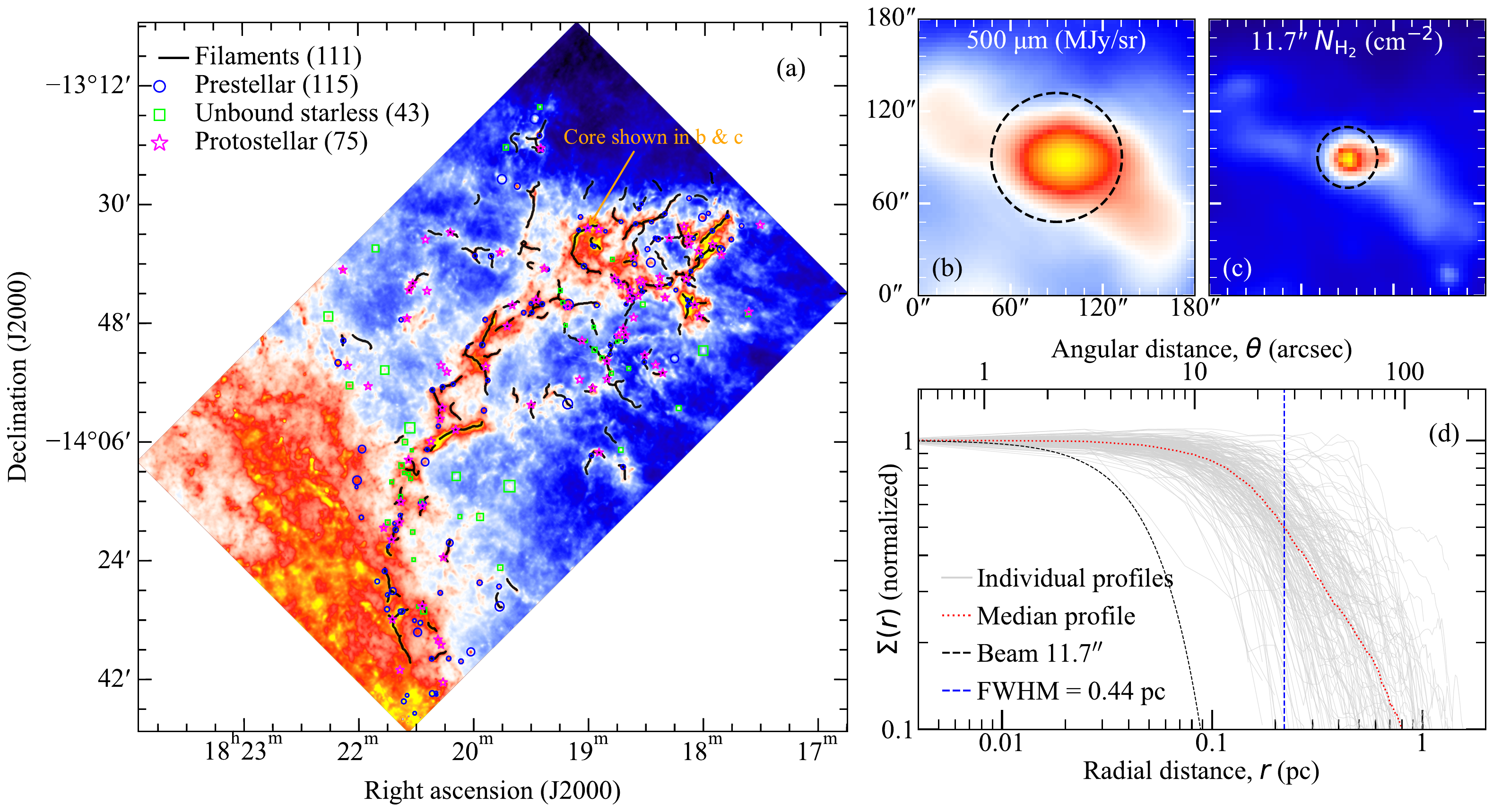}
    \caption{
    Dense cores and filaments in M16. \textit{Left:} High-resolution surface density map showing the spatial distribution of 233 dense cores (colored symbols) and skeletons of 111 extracted filaments (curves). \textit{Top right:} Zoom-in maps of a representative dense core in 500\,$\mu$m intensity (36.3$^{\prime\prime}$ resolution) and surface density (11.7$^{\prime\prime}$ resolution), illustrating the detailed structure and surrounding filamentary environment. \textit{Bottom right:} Composite radial profiles of filaments. Gray curves show individual one-sided profiles, the red dotted curve shows the median profile, the black dashed curve represents the beam profile, and the blue dashed line indicates the median filament width (0.44 pc).
    }
    \label{fig:cores_filaments}
\end{figure*}

Figure~\ref{fig:cores_filaments}d shows composite radial profiles from 196 individual filament sides (100 left sides and 96 right sides). The median profile shows a smooth decrease from the center, with a half-maximum width of 0.44 pc, consistent with the median width from individual filament measurements (0.42 pc). The radial profiles show considerable scatter in the outer regions, reflecting the diversity of filament environments in M16.

\begin{figure*}
    \centering
    \includegraphics[width=0.95\hsize]{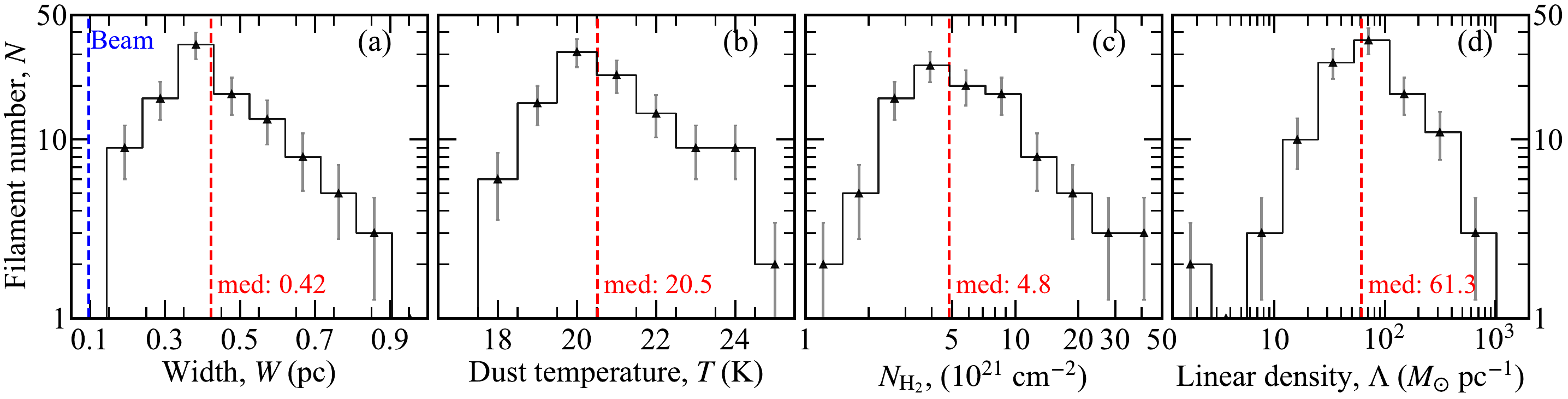}
    \caption{
    Distributions of physical properties for 111 filaments in M16. Histograms show: (\textit{a}) filament width $W$, (\textit{b}) crest temperature $T$, (\textit{c}) crest surface density $N_{\mathrm{H}_2}$, and (\textit{d}) linear mass density $\Lambda$. Red dashed lines indicate median values. The blue dashed line in panel (\textit{a}) shows the spatial resolution (11.7$^{\prime\prime}$ = 0.096 pc) for comparison with filament widths.
    }
    \label{fig:filament_properties}
\end{figure*}

Figure~\ref{fig:filament_properties} presents distributions of key physical parameters for the 111 filaments in M16, and Table~\ref{tab:filament_statistics} summarizes their statistical properties. The filament widths range from 0.16 to 1.16 pc with a median of 0.42 pc. The distribution is right-skewed (skewness $\gamma=1.2$) with a coefficient of variation (CV) of 41\%, indicating moderate dispersion. Dust temperatures along filament crests range from 17.6 to 25.5 K with a median of 20.5 K. The distribution is relatively symmetric ($\gamma=0.6$) with low dispersion (CV = 8.3\%), suggesting uniform thermal conditions along the filament network.

Crest surface densities span a wide dynamic range of $0.3{-}43.9\times 10^{21}$ cm$^{-2}$, with a median of $4.9\times 10^{21}$ cm$^{-2}$. The distribution is strongly right-skewed ($\gamma=2.7$) with high dispersion (CV = 102\%), reflecting the presence of both diffuse and dense filaments in the M16 region. Linear densities range from 2.1 to 786 $M_\odot$ pc$^{-1}$ with a median of 61.3 $M_\odot$ pc$^{-1}$. This parameter also exhibits strong right skewness ($\gamma=2.9$) and high dispersion (CV = 124\%), consistent with the wide range of filament masses in the region.

The complete catalog of all 111 filaments, including their physical properties and spatial parameters, is provided in Appendix~\ref{app:filament_catalog}, with the full version available online.

\begin{table}
\centering
\caption{Statistical properties of filaments in M16.}
\label{tab:filament_statistics}
\begin{tabular}{lccc}
\toprule
Parameter & Mean $\pm$ stdev & Median & Range \\
\midrule
$W$ (pc) & $0.47 \pm 0.20$ & 0.42 & 0.16--1.16 \\
$T$ (K) & $20.9 \pm 1.7$ & 20.5 & 17.6--25.5 \\
$N_{\mathrm{H}_2}$ ($10^{21}$ cm$^{-2}$) & $7.5 \pm 8$ & 5 & 0.28--44 \\
$\Lambda$ ($M_\odot$ pc$^{-1}$) & $105 \pm 130$ & 61 & 2.1--790 \\
\bottomrule
\end{tabular}
\end{table}

\subsection{Analysis of the shell structure}
\label{sect:radial_analysis}

To investigate how stellar feedback influences structure formation in M16, we analyzed a prominent arc-like segment of the shell driven by the central NGC 6611 cluster. This arc has a radius of curvature $R_{\rm s} \approx 25$ pc, subtends $110^\circ$, and represents material swept up by stellar winds and radiation pressure. We excluded the Galactic plane region due to contamination.

\begin{figure*}
    \centering
    \includegraphics[width=\hsize]{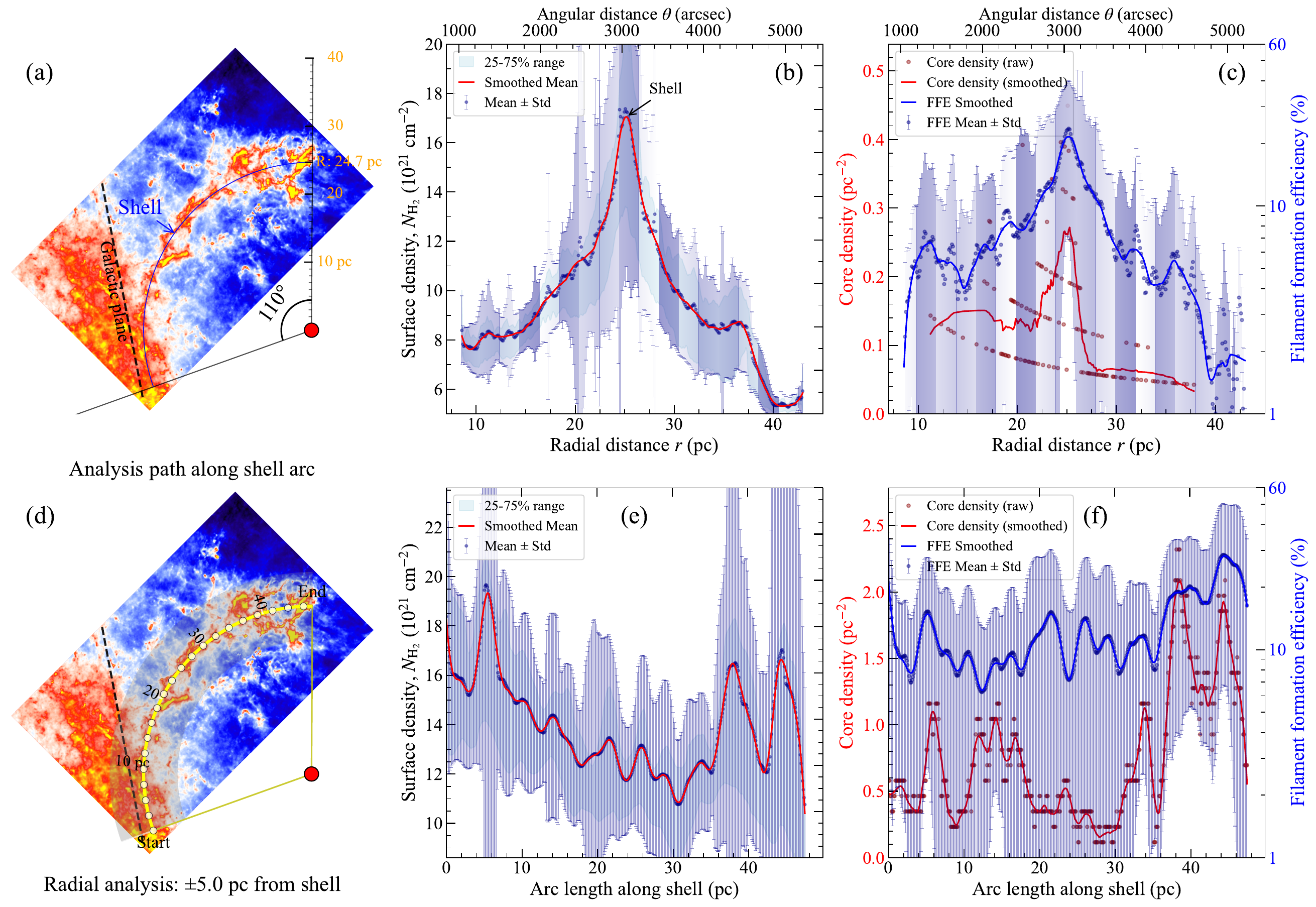}
\caption{
Radial analysis of the large-scale shell structure in M16. 
\textit{Left:} High-resolution surface density map with the shell boundary (blue dashed arc at $R_{\rm s}=24.7$ pc) and Galactic plane exclusion region (black dashed line). 
\textit{Middle:} Radial profiles of H$_{2}$ surface density (\textit{top}) and azimuthal variations along the shell arc (\textit{bottom}). 
\textit{Right:} Radial profiles of core number density (\textit{top}) and filament formation efficiency (FFE, \textit{bottom}). 
Shaded regions represent the 25--75th percentile range and red/blue curves show smoothed trends. 
The horizontal striations visible in the core density scatter plot (dark red points) are a result of the quantization of core counts (integer values) within fixed-width annuli of $\sim$0.1\,pc, leading to discrete values of cores per unit area.} 
The arrow marks a prominent density enhancement associated with shell compression. 
Both core density and FFE exhibit strong peaks coincident with the shell radius, demonstrating enhanced structure formation in the compressed layer.
    \label{fig:shell_analysis}
\end{figure*}

We calculated radial profiles of three key quantities to characterize structure formation within the shell: (1) H$_2$ surface density $N_{\mathrm{H}_2}$ (cm$^{-2}$), (2) core density (number of dense cores per unit area, cores pc$^{-2}$), and (3) filament formation efficiency (FFE, the mass fraction of gas in filamentary structures, expressed as a percentage). The FFE is computed as the ratio of filament mass surface density to total gas surface density. We analyzed the shell along a 110$^\circ$ arc sector, averaging radially over $\pm 5.0$ pc from the shell radius ($R_{\rm s} = 24.7$ pc). Results are presented in Fig.~\ref{fig:shell_analysis}.

Compared to the overall cloud average, the core surface density is enhanced by a factor of 1.5 within the shell region, while the FFE shows an even stronger enhancement of 2.3. Furthermore, both the core surface density and the FFE reach their peak values of 0.27 cores pc$^{-2}$ and 21.5\%, respectively, at $r \approx 25$ pc, closely aligned with the shell radius $R_{\rm s}$. 
We note that the discrete horizontal striations observed in the core density scatter plot are statistical in nature, arising from the conversion of small integer core counts into areal densities within annular bins of constant width.
These density enhancements and their close spatial alignment suggest a coupled formation sequence in which shell compression promotes filament formation and their subsequent fragmentation into cores.

The Pearson correlation coefficient between core density and FFE across the entire analyzed region is $r=0.67$ ($p = 2.6 \times 10^{-14}$), indicating a statistically significant moderate positive correlation. This suggests that core formation and filament formation are physically coupled processes in M16. These quantitative comparisons between the shell region and the entire cloud are summarized in Table~\ref{tab:core_density_ffe_stats}.

All parameters reveal substantial variations along the shell arc (Fig.~\ref{fig:shell_analysis}e,f). The H$_2$ surface density and core surface density vary by 64 and 291\%, respectively, along the arc, while the FFE varies by 176\%. These variations indicate inhomogeneous structure formation within the shell, likely reflecting the complex interplay between the expanding shell and pre-existing cloud structure.

\subsubsection{Ambient density and dynamical age}
\label{ssec:dynamics}

The dynamical evolution of the wind-blown bubble depends critically on the initial density $\rho_0$ of the unperturbed medium. We estimate $\rho_0$ using two independent approaches that bracket the ambient density.

We use the background H$_2$ surface density at $r > 30$ pc (outside the bubble's influence), where $N_{\mathrm{H}_2}^{\rm bg} \approx 8 \times 10^{21}$ cm$^{-2}$ (Fig.~\ref{fig:m16_maps}). Converting to volume density requires an estimate of the cloud depth. The radial profile of the shell (Fig. 6b) shows a half-maximum width of $\sim$10 pc (between radii $\sim$20 and 30 pc), extending to widths of $\sim$20 pc, typical of molecular cloud complexes. Adopting this characteristic depth scale $L \sim 10$ pc yields an initial number density $n_{0} \approx N_{\mathrm{H}_2}^{\rm bg}/L \sim 260$ cm$^{-3}$ and mass density $\rho_0 = \mu m_{\rm H} n_{0} \sim 1.2 \times 10^{-21}$\,g\,cm$^{-3}$.

Alternatively, we can estimate $\rho_0$ from the total mass within the bubble ($r < 30$ pc). Integrating the surface density map over this region and extrapolating the analyzed 110$^\circ$ arc to the full shell (multiplying by $3.27$) yields a total mass $7.6\times 10^{5}$\,$M_{\odot}$ of the bubble. Dividing by the swept volume gives $\rho_0 \sim 4.5 \times 10^{-22}$ g cm$^{-3}$, approximately a factor of 2.7 lower than in the first method. This lower value may reflect mass loss through the shell or incomplete sampling of the swept material.

The mechanical luminosity from NGC 6611's 13--15 O-type stars
\citep{Evans+2005, Sana+2009, Stoop+2023} is estimated as
$L_{\rm w} \approx 2.0 \times 10^{38}$\,erg\,s$^{-1}$, based on typical
O-star mass-loss rates ($\dot{M} \sim 5 \times 10^{-6}$\,$M_\odot$\,yr$^{-1}$)
and wind velocities ($v_{\infty} \sim 3000$ km s$^{-1}$) \citep{Vink+2000}.

Using the standard wind-blown bubble model \citep{Weaver+1977}, the shell radius evolves as
\begin{equation}
R_{\rm s}(t) = 0.76 \left( \frac{L_{\rm w}}{\rho_0} \right)^{1/5} t^{3/5}.
\label{eq:bubble_radius_main}
\end{equation}
For the observed $R_{\rm s} \approx 25$ pc and our updated $L_{\rm w}$, the two density estimates yield dynamical time $t_{\rm d} \sim 1.3$ and 1.0\,Myr, and shell expansion velocity $v_{\rm s} = (3/5)R_{\rm s}/t_{\rm d}$ of $\sim 12$ and 16\,km\,s$^{-1}$, respectively, consistent with the age of NGC 6611's youngest stellar population ($\sim$1--2 Myr; \citealt{Hillenbrand+1993}) and observational velocity constraints \citep{Flagey+2011}. A complementary analysis of M16 shell dynamics by \citet{Karim+2025} also finds evidence for an expanding structure and estimates a kinematic age of $\sim$2\,Myr from the present-day radius and a constant $10$\,km\,s$^{-1}$ expansion velocity. The modest difference with our age range arises because we adopt a wind-blown bubble evolution model that accounts for deceleration \citep{Weaver+1977}, whereas \citet{Karim+2025} assume a linear expansion. Both estimates are consistent with the cluster age and support ongoing shell fragmentation.

\begin{table*}
\centering
\caption{Statistical comparisons of the shell and entire M16 cloud (excluding Galactic plane).}
\label{tab:core_density_ffe_stats}
\begin{tabular}{lccc}
\toprule
Parameter & Entire region (7--47 pc) & Shell region ($24.7\pm 2.5$ pc) & Enhancement (shell/overall) \\
\midrule
Density of cores (cores\,pc$^{-2}$) & & & \\
~~Mean & $0.13 \pm 0.09$ & $0.20 \pm 0.10$ & 1.5 \\
~~Median & 0.10 & 0.19 & \\
~~Peak (radius) & -- & 0.27 (25.3 pc) & \\
\midrule
FFE (\%) & & & \\
~~Mean & $7.0 \pm 4.6$ & $15.9 \pm 3.9$ & 2.3 \\
~~Median & 5.9 & 15.4 & \\
~~Peak (radius) & -- & 21.5 (25.2 pc) & \\
\midrule
Density of cores to FFE ratio & 0.019 & 0.012 & -- \\
\midrule
Density of cores and FFE correlation & \multicolumn{3}{c}{$r = 0.67$ ($p = 2.5 \times 10^{-14}$)} \\
\bottomrule
\end{tabular}
\end{table*}

\section{Discussion}
\label{sect:discussion}

\subsection{Physical properties of dense cores in M16}

Our analysis of 233 reliable cores in M16 reveals several important trends in their physical properties. The core mass distribution (Fig.~\ref{fig:core_properties}b) shows a median mass of 5.6\,$M_\odot$ with a long tail extending to 231\,$M_\odot$, indicating the presence of both low-mass and high-mass cores. The temperature distribution (Fig.~\ref{fig:core_properties}a) spans a relatively narrow range of 16--27\,K with a median of 20\,K, suggesting that most cores are moderately shielded from external radiation.

The $\alpha_{\text{BE}}$ distribution (Fig.~\ref{fig:core_properties}c) clearly distinguishes between core types:
prestellar cores ($\alpha_{\text{BE}}\leq 2$) have a median $\alpha_{\text{BE}}$ of 1.01, while unbound cores have a median of 2.68. The distribution of average H$_{2}$ number density in cores (Fig.~\ref{fig:core_properties}d) spans three orders of magnitude from $4.5\times10^2$ to $1.1\times10^5$\,cm$^{-3}$, with a median of $4.8\times10^3$\,cm$^{-3}$.

The mass-size relation (Fig.~\ref{fig:mass_size}) shows that prestellar cores tend to occupy a distinct region in parameter space, generally having higher masses for a given size compared to unbound cores. Several prestellar and protostellar cores lie above the empirical threshold for massive star formation proposed by \citet{Kauffmann+2010}, suggesting that M16 is in the process of forming massive stars.

Our results reveal distinct physical properties between bound and unbound cores (Table~\ref{tab:core_statistics}). Prestellar and protostellar cores are significantly more massive, with mean masses of approximately 13\,$M_\odot$, while unbound cores have mean masses of only 2.9\,$M_\odot$. The mean $\alpha_{\text{BE}}$ values also clearly separate the core types: approximately 1.0 for prestellar cores, 3.2 for unbound cores, and 1.3 for protostellar cores. Although 21\% of protostellar cores have $\alpha_{\text{BE}} > 2$, this does not necessarily indicate that these cores are gravitationally unbound, but rather reflects either underestimated masses (due to observational uncertainties) or the approximate nature of the Bonnor-Ebert boundedness criterion when applied to cores with internal heating from protostars.

The power-law fits reported in Section~\ref{sec:core_selection_classification} reveal distinct scaling behaviors among the three core populations. Prestellar cores exhibit a shallow slope ($\gamma = 0.69$), indicating that their surface density decreases with increasing radius — a configuration where the central region is more concentrated while the outer envelope remains relatively diffuse. Unbound cores show a near-linear relation ($\gamma \approx 1.2$), implying approximately constant surface density, characteristic of uniform-density clumps that have not undergone significant gravitational contraction. Protostellar cores have a steep slope ($\gamma = 2.5$), reflecting both the presence of dense central protostars and efficient accretion of surrounding material, combined with a selection bias towards bright, compact sources in the 70~$\mu$m band. A seemingly puzzling observation is that prestellar cores ($\gamma = 0.69$) have a flatter slope than unbound cores ($\gamma = 1.24$), yet they are gravitationally bound ($\alpha_{\rm BE}\leq 2$) while unbound cores are not ($\alpha_{\rm BE}>2$). This apparent discrepancy is resolved by examining the definition of $\alpha_{\rm BE}$. From $M_{\rm BE} = 2.4 R_{\rm c} c_{\rm s}^2 / G$ and $c_{\rm s}^2 \propto T$, we have $\alpha_{\rm BE} = M_{\rm BE}/M_{\rm c} \propto R_{\rm c} T / M_{\rm c}$. Because the temperature varies only weakly among core populations (Table~\ref{tab:core_statistics}), $\alpha_{\rm BE}$ is primarily governed by the mass-to-radius ratio $M_{\rm c}/R_{\rm c}$. Prestellar cores have a median mass of $7\,M_\odot$ and a median radius of $0.18$ pc, yielding $M_{\rm c}/R_{\rm c} \approx 39\,M_\odot$ pc$^{-1}$, whereas unbound cores have $2.6\,M_\odot$ and $0.16$ pc, giving $M_{\rm c}/R_{\rm c} \approx 16\,M_\odot$ pc$^{-1}$. The factor of $\sim2.4$ higher mass-to-radius ratio of prestellar cores compensates for their flatter $M$–$R$ scaling exponent, rendering them gravitationally bound. Hence $\alpha_{\rm BE}$ reflects the absolute mass concentration, not the slope of the mass–radius relation. The low $R^2$ value for prestellar cores ($0.07$) indicates a large scatter in their mass–radius distribution, likely due to a wide range of evolutionary stages within this population — from recently assembled, still-loose condensations to more advanced, centrally concentrated cores.

\subsection{Impact of non-thermal motions on core stability}
\label{sect:nonthermal}

The Bonnor-Ebert (BE) analysis presented in Section~\ref{sec:core_selection_classification} and the $\alpha_{\mathrm{BE}}$ distribution shown in Figure~\ref{fig:core_properties}c provide a useful first-order classification of gravitational boundedness based solely on thermal pressure support. However, it is well established that molecular cloud cores exhibit significant non-thermal (turbulent) motions that contribute additional support against gravitational collapse \citep[e.g.,][]{MacLow+Ossenkopf2000, Padoan+2001, Li+2013}.

To evaluate the influence of turbulent support, we compute an effective sound speed $c_{\mathrm{eff}} = \sqrt{c_{\mathrm{s}}^2 + \sigma_{\mathrm{NT}}^2}$, where $\sigma_{\mathrm{NT}}$ is the non-thermal velocity dispersion. In the specific case of M16, APEX CO\,(3--2) observations reveal a typical line width (FWHM) of $\sim 2$ km s$^{-1}$ \citep{Karim+2025}, from which the non-thermal velocity dispersion is derived as $\sigma_{\mathrm{NT}} = \mathrm{FWHM} / (2\sqrt{2\ln 2}) \approx 0.85$ km s$^{-1}$. 
We caution that CO\,(3--2) traces relatively warm, collisionally excited gas and may overestimate the non-thermal line width of the cold, dense cores traced by the \textit{Herschel} dust emission \citep[e.g.,][]{Dumke+2001, Johnstone+2010}. 
This yields an effective sound speed $c_{\mathrm{eff}} \approx 0.89$ km s$^{-1}$, approximately three times the thermal sound speed $c_{\mathrm{s}} \approx 0.27$ km s$^{-1}$ for $T=20$ K. Adopting a turbulent Bonnor-Ebert mass $M_{\mathrm{BE,turb}} = 2.4 R_{\mathrm{c}} c_{\mathrm{eff}}^2 / G$ raises the stability threshold by nearly an order of magnitude and would formally reclassify the majority of our cores as unbound. This underscores the fact that the classical BE criterion is best viewed not as an absolute classifier, but rather as a first-order comparative tool for assessing the relative state of gas concentrations, and that virial analysis including external pressure provides a more reliable assessment \citep{Pattle+2025}.

The inclusion of non-thermal (turbulent) support does not alter the principal conclusions of this work.
While adopting an effective sound speed \(c_{\mathrm{eff}} \approx 0.89\)\,km\,s\(^{-1}\) (Sect.~\ref{sect:nonthermal}) raises the stability threshold and would formally reclassify many individual cores as unbound, our analysis of feedback-driven structure formation deliberately employs the full, unbiased sample of dense cores and filaments. 
The observed factor of \(\sim1.5\) enhancement in core surface density and the factor of \(\sim2.3\) enhancement in filament formation efficiency within the shell (Sect.~\ref{sect:radial_analysis}) are statistical measurements of the total local output of the fragmentation process, independent of the internal dynamical state assigned to any particular core.
Applying a stability-dependent filter—whether based on thermal pressure alone or on the combined thermal and turbulent support—would merely introduce a model-dependent selection bias without changing the fundamental observational result: the expanding shell has significantly elevated the condensation rate of dense substructures from the ambient molecular medium.

\subsection{Filamentary structures in M16}

The systematic measurement of 111 filaments in M16 provides important insights into the filamentary network of this massive star-forming region. Our analysis reveals several key characteristics that distinguish M16 filaments from those observed in nearby low-mass star-forming regions.

\subsubsection{Filament widths and resolution considerations}

Studies of nearby molecular clouds observed with \textit{Herschel} have suggested that filament widths are largely independent of surface density, typically measuring $\sim$0.1 pc \citep{Arzoumanian+2011, Arzoumanian+2019}. Whether this characteristic width is universal -- particularly in high-mass star-forming environments -- remains actively debated \citep{Panopoulou+2022, Andre+2022}. 

In our sample, filament widths range from 0.16 to 1.16 pc with a median of 0.42 pc -- significantly larger than the characteristic width reported for nearby clouds like Taurus and Ophiuchus \citep[e.g.,][]{Li+2012, Palmeirim+2013, Jia+2025}, but consistent with the median width of 0.3 pc found in the Vela C molecular cloud \citep{Li+2023}. Given the angular resolution of our surface density map ($11.7^{\prime\prime}$) and the adopted distance of 1700 pc to M16, this corresponds to a spatial resolution of approximately 0.096 pc. While all filaments satisfy traditional resolution criteria (measured width $\gg$ beam size), recent work by \cite{MenshchikovZhang2026subm} demonstrated that filaments with shallow radial profiles remain effectively unresolved even when their measured width greatly exceeds the beam size. Therefore, some of our filaments -- particularly those with shallower profiles -- may not be fully resolved despite meeting conventional resolution criteria.

The wide range of observed widths (0.16--1.16 pc) and right-skewed distribution (Fig.~\ref{fig:filament_properties}a) suggest a diversity of filament formation mechanisms and evolutionary states. While most filaments cluster around 0.4 pc, a significant population of broader structures exists. This enhanced median width compared to low-mass regions may reflect several factors: (1) the influence of stronger turbulence, magnetic fields, and stellar feedback in the high-mass star-forming environment \citep{Ren+2023}; (2) different evolutionary stages or mass loading; or (3) external compression and widening by the expanding shell. The broader filaments may represent either younger structures that have not yet collapsed to a characteristic width or filaments actively shaped by environmental forces.

\subsubsection{Temperature and surface density distributions}

The dust temperatures along filament crests span 17.6--25.5 K (median 20.5 K), systematically higher than typical temperatures in nearby star-forming regions (10--15 K). This elevation is consistent with intense heating from the central NGC 6611 cluster. The relatively narrow temperature range suggests uniform heating across the filament network, likely due to efficient penetration of radiation through the dense filamentary structures.

Surface densities along filament crests exhibit a very wide dynamic range ($0.28{-}43.9\times 10^{21}$ cm$^{-2}$), spanning nearly two orders of magnitude. The strongly right-skewed distribution indicates that while most filaments have moderate surface densities (median $4.85\times 10^{21}$ cm$^{-2}$), a small population of extremely dense filaments exists. These high-density structures may be critical sites for massive star formation, providing the necessary reservoir of dense gas for high-mass core assembly.

\subsubsection{Gravitational stability and linear densities}

Linear densities in our sample range from 2.1 to 786\,$M_\odot$ pc$^{-1}$ (median 61\,$M_\odot$ pc$^{-1}$), showing substantial variation across the filament population. Classical theoretical models \citep{Stod1963, Ostriker1964} predict a critical linear density $\Lambda_{\text{c}}=2c_{\rm s}^2/G$ above which infinitely long, isothermal, hydrostatic, non-magnetic cylindrical filaments become gravitationally unstable and fragment. For the typical filament temperature of 20 K in M16, $\Lambda_{\text{c}}\approx 32$ $M_\odot$ pc$^{-1}$. Remarkably, 76\% of M16 filaments exceed this critical value, with a median linear density of $1.9\Lambda_{\text{c}}$, suggesting widespread gravitational instability and efficient fragmentation into cores. This may explain the high core formation efficiency observed in M16.

However, applying this instability criterion to actual \textit{Herschel} observations requires significant caution. The observed filaments (Figs.~\ref{fig:m16_maps}, \ref{fig:cores_filaments}) systematically deviate from idealized model assumptions: they exhibit finite lengths, non-isothermal conditions, detectable magnetic fields, and no clear evidence of hydrostatic equilibrium. These structures exist within complex, inhomogeneous environments, displaying pronounced curvature, interactions with neighboring features, radial temperature gradients, and substantial variations in both width and linear density along their lengths. Consequently, the theoretical $\Lambda_{\text{c}}$ should be regarded only as an approximate, order-of-magnitude indicator of potential gravitational instability rather than a precise threshold.

\subsubsection{Radial profiles and environmental influences}

The radial profiles (Fig.~\ref{fig:cores_filaments}d) reveal that M16 filaments exhibit relatively shallow outer profiles, with the median profile showing a gradual decline from the crest. This suggests that M16 filaments are embedded in complex, structured environments. The considerable scatter in the composite profile's outer regions reflects both the diversity of filament environments and the spatially variable influence of the expanding shell.

Preliminary analysis indicates systematic variations in filament properties with position relative to the shell. Filaments within the shell region tend to be wider and exhibit higher surface densities than those outside, consistent with compression by the expanding bubble. This spatial dependence suggests that stellar feedback plays an active role in shaping the filamentary network and may enhance gravitational instability in compressed regions.

\subsubsection{Comparison with other filament extraction algorithms}
\label{ssec:algo_comparison}

Previous \textit{Herschel} studies of M16 used the \textit{disperse} algorithm 
\citep{Sousbie2011, Hill+2012} to trace continuous, branching filamentary networks. 
Understanding why \textit{getsf} and \textit{disperse} produce different results 
requires examining how each algorithm defines filament skeletons, because the 
skeleton is not merely a visualization tool: it defines the orthogonal directions 
(normals) along which radial profiles are extracted, and therefore determines all 
derived physical quantities -- widths, linear densities, and masses. \textit{Disperse} 
traces skeletons directly in the observed image, where fluctuations on all spatial 
scales are present. Small-scale noise, large-scale background structure, and
embedded sources all distort the local intensity ridge, causing the skeleton to deviate 
from the true filament crest. The resulting normals diverge from the correct orthogonal 
directions, so that radial profiles are measured along oblique cuts, biasing widths 
and linear densities. This problem is most severe for resolved filaments, where 
small-scale fluctuations are large relative to the crest curvature. The M16 filaments, 
with a median width of $\sim$0.42\,pc ($\sim$51$^{\prime\prime}$ at 1700\,pc), 
are very well resolved with respect to the $11.7^{\prime\prime}$ map resolution, 
placing them squarely in the regime where this bias is strongest. Quantitative tests 
using simulated straight filaments confirm this: even at a crest-to-noise ratio of 20, 
\textit{disperse} produces markedly wiggly skeletons, whereas \textit{getsf} 
recovers a straight crest on all spatial scales \citep{Zhang+2026subm}.
This is why \cite{Hill+2012} did not attempt to measure filament widths or linear 
densities from \textit{disperse} skeletons in M16.

The \textit{getsf} method avoids these biases through a multiscale approach. It 
operates on spatially decomposed single-scale images in which contributions from 
scales outside a narrow range are removed; it extracts and subtracts sources before 
deriving skeletons; and it detects skeletons within a controlled range of scales, 
suppressing both small-scale noise and large-scale background undulations. 
The result is that \textit{getsf} skeletons closely follow the true filament crests even 
for wide, low-contrast structures, yielding reliable normals and therefore reliable 
profile measurements. This is why \textit{getsf} identifies 111 discrete, non-branching 
filaments in M16, while \textit{disperse} yields a more spatially continuous network: 
\textit{disperse} stitches together irregular skeleton segments that happen to run 
along intensity ridges in the noisy observed image, whereas \textit{getsf} identifies 
only features whose crests are robustly recovered after scale decomposition and 
source removal. The caveat readers should keep in mind is that \textit{getsf}'s 
catalog may be more conservative in total filamentary extent, potentially missing 
diffuse, low-contrast inter-core bridges. However, for measuring local physical properties 
and performing spatial comparisons between the shell interior and exterior, accurate 
skeletons are a prerequisite, and our main conclusions -- the enhanced FFE and core 
density within the shell, the supercritical nature of most filaments, and the hierarchical 
fragmentation sequence -- are robust to algorithm choice because they depend on those 
local measurements, not on skeleton connectivity.

\subsection{The large-scale shell in M16 as a site of enhanced structure formation}

Our radial analysis confirms that the $\sim$60 pc diameter shell in M16, driven by stellar feedback from the central NGC 6611 cluster, represents a significant site of enhanced structure formation. The more than two-fold enhancement of filament formation efficiency (FFE) within the shell strongly supports theoretical frameworks of shell fragmentation and ``collect-and-collapse'' processes \citep{Elmegreen+Lada1977}.

\subsubsection{Physical conditions for shell fragmentation}

The physical conditions for gravitational fragmentation of a swept-up shell can be evaluated using the criterion derived by \citet{Elmegreen+1994, Wunsch+Palous2001}:
\begin{equation}
\Sigma_{\text{c}} \approx \frac{c_{\rm s}^2}{G R_{\rm s}} \left(1+\left(\frac{v_{\rm s}}{c_{\rm s}}\right)^{2}\right)^{1/2},
\label{eq:fragmentation_criterion}
\end{equation}
where $\Sigma_{\text{c}}$ is the critical surface density for fragmentation, $c_{\rm s}$ is the isothermal sound speed, $G$ is the gravitational constant, $R_{\rm s}$ is the shell radius, and $v_{\rm s}$ is the expansion velocity. This criterion balances the gravitational instability of the compressed layer against thermal and kinetic support.

For the M16 shell with $v_{\rm s} = 12$--16\,km\,s$^{-1}$ (from our dynamical analysis), and $c_{\rm s} \approx 0.3$ km s$^{-1}$ (corresponding to $T \approx 20$ K), Eq.~(\ref{eq:fragmentation_criterion}) yields $\Sigma_{\text{c}} \sim 1 \times 10^{21}$ cm$^{-2}$. The observed surface densities within the shell region are substantially higher, reaching $\sim 8 \times 10^{21}$ cm$^{-2}$ (Fig.~\ref{fig:m16_maps}), approximately 8 times the critical threshold. This significant excess provides strong quantitative support for ongoing gravitational fragmentation and explains the enhanced structure formation observed in the shell.

The characteristic timescale for shell fragmentation can be estimated as
\begin{equation}
t_{\text{f}} = \frac{R_{\rm s}}{v_{\rm s}} \sim 1.5{-}2.0 \times 10^6 \text{ yr},
\label{eq:fragmentation_timescale}
\end{equation}
representing the time required for perturbations to grow and the shell to fragment. This is comparable to the estimated dynamical age of the M16 shell (1.0--1.3 Myr from Sect.~\ref{ssec:dynamics}), indicating that we are observing the shell during or shortly after the onset of gravitational fragmentation. The fact that $t_{\rm d}\sim t_{\rm f}$ suggests active, ongoing fragmentation rather than a completed process, consistent with the observed enhancements in both filament formation efficiency ($\times2.3$) and core density $\times1.5$) within the shell region. This temporal coincidence strengthens the interpretation that the observed filaments and cores are products of feedback-driven fragmentation rather than pre-existing structures swept up by the shell.

\subsubsection{Shell dynamics and compression mechanism}

The evolution of the expanding shell is governed by the equation of motion balancing the driving pressure from stellar feedback with the inertia of the swept-up ambient medium \citep{Weaver+1977}:
\begin{equation}
\frac{{\rm d}}{{\rm d}t} \left( M_{\rm s} \frac{{\rm d}R_{\rm s}}{{\rm d}t} \right) = 4\pi R_{\rm s}^2 (P_{\text{in}} - P_0),
\label{eq:shell_motion}
\end{equation}
where $M_{\rm s}$ is the swept-up mass, $P_{\text{in}}$ is the interior pressure maintained by stellar winds and radiation from NGC 6611, and $P_0$ is the ambient pressure of the surrounding molecular cloud.

This dynamical framework describes how stellar winds and radiation pressure from the central cluster continuously sweep up and compress surrounding material into a dense shell. As the shell expands according to Eq.~(\ref{eq:shell_motion}), material accumulates in a thin, compressed layer. When the local surface density of this layer exceeds the critical value for gravitational instability (Eq.~\ref{eq:fragmentation_criterion}), it fragments into the filamentary structures observed in our analysis.

This physical picture is corroborated by observations and numerical simulations of M16 by \citet{Tremblin+2013}, who demonstrated that compressed layers at the edges of {\HII} regions are preferential sites for the formation of pillar-like structures and globules. The spatial coincidence between the shell boundary and enhanced filament formation in our data provides direct observational evidence for this compression-driven fragmentation mechanism.

\subsubsection{Hierarchical fragmentation: from shell to cores}

Beyond filament formation, our analysis reveals a hierarchical fragmentation sequence operating within the shell. The core number density is enhanced by a factor of 1.5 within the shell compared to the exterior, with the spatial distribution of cores closely aligned with FFE peaks. The moderate correlation between core density and FFE ($r = 0.67$) suggests a causative relationship: compressed gas first fragments into filaments, which subsequently undergo secondary fragmentation into dense cores through gravitational instabilities.

This ``shell $\rightarrow$ filaments $\rightarrow$ cores'' hierarchical fragmentation sequence represents a multi-scale cascade of gravitational instability. The initial compression by the expanding shell creates conditions favorable for large-scale fragmentation into filaments (Eq.~\ref{eq:fragmentation_criterion}). Once formed, these filaments -- many of which are supercritical with respect to their linear mass density (Sect.~\ref{sect:filament_measurement}) -- undergo further fragmentation into cores. This hierarchical process is consistent with structural hierarchies observed in other star-forming regions \citep[e.g.,][]{ZhangC+2020} and provides a coherent framework for understanding the enhanced star formation activity within the M16 shell.

The moderate rather than strong correlation between core density and FFE likely reflects the varying evolutionary states of filaments within the shell. Younger filaments may have formed recently through shell compression but not yet fragmented into cores, while more evolved filaments have already produced substantial core populations. This temporal diversity within the shell population naturally produces scatter in the core density--FFE relationship while maintaining an overall positive correlation.

\subsubsection{Implications for triggered star formation}

The quantitative agreement between theoretical predictions and observed properties -- including critical surface densities (Eq.~\ref{eq:fragmentation_criterion}), fragmentation timescales (Eq.~\ref{eq:fragmentation_timescale}), and spatial distributions -- provides evidence that stellar feedback from NGC 6611 has actively triggered structure formation in M16. The shell does not merely redistribute pre-existing material but creates physical conditions conducive to enhanced gravitational instability and subsequent star formation.

The remarkably high FFE within the shell ($\sim$16\%, representing a 2.3-fold enhancement over the exterior) demonstrates that feedback-driven compression efficiently reorganizes diffuse material into potential star-forming sites. The measured filament properties strongly support this triggered formation scenario. Filaments within the shell exhibit clear signatures of compression -- broader widths and elevated surface densities compared to exterior filaments -- and are mostly supercritical (76\% exceed the critical linear density for fragmentation). This supercritical state makes them prone to secondary fragmentation into cores, as evidenced by the 1.5-fold enhancement in core density within the shell. The hierarchical sequence from shell compression to filament formation to core fragmentation provides a coherent physical framework linking large-scale feedback to small-scale star formation.

\citet{Flagey+2011} previously identified anomalously warm dust and enhanced radiation fields within the M16 shell, attributing these to shell dynamics (including wind shocks) or potentially a supernova remnant. This additional energy input may not only explain the observed heating but could further promote structure formation through enhanced compression and turbulent dissipation, potentially accelerating the fragmentation process beyond what stellar winds and radiation pressure alone would produce.

This represents a clear observational example of positive feedback, where massive star formation in the cluster core induces secondary structure formation in the surrounding molecular cloud. The efficiency of this mechanism suggests that feedback-driven compression may be a dominant pathway for star formation in massive star-forming regions, potentially explaining the high star formation rates observed in such environments. Understanding the relative contributions of spontaneous versus triggered star formation -- and the conditions under which feedback enhances rather than suppresses star formation -- remains crucial for developing comprehensive models of clustered star formation and the self-regulation of star formation activity in giant molecular clouds.

\section{Conclusion}
\label{sect:conclusion}

We have conducted a comprehensive analysis of dense cores and filamentary structures in the M16 molecular cloud using high-resolution ($11.7^{\prime\prime}$) surface density and temperature maps derived from \textit{Herschel} observations. Using the \textit{hires} algorithm for map construction and the \textit{getsf} method for source extraction \citep{Menshchikov2021b}, we identified 233 reliable cores and 111 well-defined filaments.

The core population comprises 75 protostellar, 115 prestellar, and 43 unbound cores, with median properties of 5.6\,$M_\odot$ and 20.1\,K. Approximately 73\% of starless cores are potentially gravitationally bound based on their virial parameters. Prestellar cores are systematically more massive (median 6.8\,$M_\odot$) than protostellar (5.5\,$M_\odot$) or unbound cores (2.6\,$M_\odot$), consistent with early-stage gravitational concentration.

The filament population exhibits a median width of 0.4\,pc -- significantly larger than the characteristic $\sim$0.1\,pc observed in nearby low-mass star-forming regions -- and a median linear density of 61\,$M_\odot$\,pc$^{-1}$. Remarkably, 76\% of filaments are supercritical with respect to gravitational instability, indicating widespread conditions favorable for core formation through filament fragmentation.

Our radial analysis of the $\sim$60 pc diameter shell driven by the central NGC 6611 cluster reveals strong enhancements in structure formation within the shell region. The filament formation efficiency (FFE) is 2.3 times higher within the shell compared to the cloud average, peaking at 21.5\% near the shell boundary ($R_{\rm s} \approx 25$ pc). Core density shows a concurrent 1.5-fold enhancement at the same location. The moderate correlation between core density and FFE ($r = 0.67$) indicates coupled formation processes operating across spatial scales.

Theoretical analysis demonstrates that the observed surface densities within the shell ($\sim8\times 10^{21}$\,cm$^{-2}$) exceed the critical threshold for gravitational fragmentation by approximately a factor of 8, indicating strongly supercritical conditions. The fragmentation timescale ($\sim$1.5--2.0 Myr) is comparable to the shell's dynamical age ($\sim$1.0--1.3 Myr), indicating we are observing the shell during an active phase of fragmentation. The extremely high FFE within the shell ($\sim$16\%) demonstrates that stellar feedback efficiently reorganizes diffuse material into filamentary structures that subsequently fragment into cores.

These findings reveal a hierarchical fragmentation sequence -- shell compression → filament formation → core formation -- operating in M16, with the expanding shell serving as a catalyst for enhanced star formation. The $\sim$60 pc shell structure provides a clear observational example of positive feedback, where massive star formation in the cluster core triggers secondary structure formation in the surrounding molecular cloud. This work demonstrates that stellar feedback can synchronously enhance both filament and core formation, supporting models of triggered star formation in feedback-driven environments and highlighting the complex interplay between large-scale compressive forces and multi-scale star formation efficiency in massive star-forming regions.

\section*{Acknowledgements}
We carried out this work at the China-Argentina Cooperation Station of NAOC/CAS. This work was supported by the Key Project of International Cooperation of the Ministry of Science and Science of China through grant 2010DFA02710, and by the National Natural Science Foundation of China through grants 11503035, 11573036, 11373009, 11433008, 11403040, and 11403041. G.Z. acknowledges support from the Postdoctoral Science Foundation of China (No. 2021T140672), and the National Natural Science Foundation of China (No. U2031118). 
N.P. acknowledges the support of the ANSO Scholarship for Young Talents (No. 2023ANP0330). 

\section*{Data Availability}

The data underlying this article are available in the article and in \textit{Herschel} Science Archive.

\bibliographystyle{mnras}
\bibliography{m16_filaments_cores.bib}

\appendix
\section{Catalogs of Dense Cores and Filaments}
\label{app:catalogs}

This appendix provides the complete catalogs of dense cores and filaments identified in M16. Tables~\ref{app:core_catalog_table} and \ref{app:filament_catalog_table} show representative entries; the full machine-readable catalogs are available as online supplementary material.

\subsection{Dense cores catalog}
\label{app:core_catalog}

Table~\ref{app:core_catalog_table} presents the catalog of 233 dense cores identified in M16, including their positions, physical properties, and classifications. Column descriptions are provided in the table notes.

\begin{table*}
\small
\setlength{\tabcolsep}{6pt}
\caption{Catalog of dense cores in M16.}
\label{app:core_catalog_table}
\centering
\begin{tabular}{c |c c c c c c c c c c c c}
\hline\noalign{\smallskip}
No. & RA  & Dec  & $A_{\rm F}$ & $B_{\rm F}$ & PA & $R_{\rm c}$ & $T_{\rm c}$ & $M_{\rm c}$ & $M_{\rm BE}$ & $\alpha_{\mathrm{BE}}$ & $n_{\rm c}$ (cm$^{-3}$) & Core type \\
    & (J2000, $^{\circ}$)   &  (J2000, $^{\circ}$) & $('')$ & $('')$ & ($^{\circ}$) & (pc) & (K) & $(M_{\odot})$ & $(M_{\odot})$ &  &  & \\
(1) & (2) & (3) & (4) & (5) & (6) & (7) & (8) & (9) & (10) & (11) & (12) & (13) \\
\noalign{\smallskip}\hline
\noalign{\smallskip}
1 & 275.2897 & -14.2524 & 55.5 & 48.6 & 142 & 0.214 & 20.8 & 48.5 & 8.8 & 0.18  & $2.1 \times 10^4$ & protostellar \\
2 & 274.8035 & -13.5618 & 39.6 & 39.1 & 11 & 0.162 & 17.1 & 87.7 & 5.5 & 0.06   & $8.5 \times 10^4$ & protostellar \\ 
$\cdots$ & $\cdots$ \\
76 & 275.2877 & -14.5301 & 60.9 & 48.8 & 107 & 0.225 & 20.5 & 231 & 9.1 & 0.04 & $8.4 \times 10^4$ & prestellar \\
77 & 275.2298 & -14.2581 & 44.7 & 42.0 & 106 & 0.178 & 20.7 & 15.7 & 7.3 & 0.46  & $1.2 \times 10^4$ & prestellar \\ 
$\cdots$ & $\cdots$ \\
191 & 275.2811 & -14.1799 & 28.4 & 28.3 & 34 & 0.117 & 21.4 & 1.80 & 4.9 & 2.74   & $4.7 \times 10^3$ & unbound \\  
$\cdots$ & $\cdots$ \\
\noalign{\smallskip}\hline
\end{tabular}

\medskip
\small
\begin{minipage}{\textwidth}
\setlength{\parindent}{0pt}
\setlength{\parskip}{2pt}
\textbf{Notes.} Column descriptions: (1) Core identification number; (2), (3) Centroid equatorial coordinates (J2000) of the core; (4), (5) Major and minor axes of the core footprint measured at the zero level; (6) Position angle of the major axis measured east of north; (7) Effective radius of the core boundary; (8) Dust temperature at the core peak position; (9) Core mass derived from surface density integration; (10) Critical Bonnor-Ebert mass; (11) Bonnor-Ebert mass ratio $\alpha_{\rm BE} = M_{\rm BE}/M_{\rm c}$ used to classify gravitational boundedness; (12) Volume-averaged number density; (13) Core classification: protostellar (associated with infrared source), prestellar (bound, $\alpha_{\rm BE} < 2$, no infrared counterpart), or unbound ($\alpha_{\rm BE} \geq 2$).
\end{minipage}
\end{table*}

\subsection{Filaments catalog}
\label{app:filament_catalog}

Table~\ref{app:filament_catalog_table} presents the catalog of 111 filaments identified in M16, including their temperatures, surface densities, and structural parameters. Column descriptions are provided in the table notes.

\begin{table*}
\small
\setlength{\tabcolsep}{4pt}
\caption{Catalog of filaments in M16.}
\label{app:filament_catalog_table}
\centering
\begin{tabular}{c | c c c c c c c c c c c c c}
\hline\noalign{\smallskip}
No. & RA  & Dec &  $L$ & $N_{\textrm{H}_{2}}$ & $\varsigma _{N_{\textrm{H}_{2}}}$ & $T$ & $\varsigma _{T}$ & $\Lambda^{\textrm{P}}_{\alpha}$ & $\Lambda^{\textrm{P}}_{\beta}$ &
$W_{\alpha}$ & $\varsigma_{{W}_{\alpha}}$ & $W_{\beta}$ & $\varsigma_{{W}_{\beta}}$ \\
    & (J2000, $^{\circ}$)   &  (J2000, $^{\circ}$) & (pc) & $(\textrm{cm}^{-2})$ & $(\textrm{cm}^{-2})$ & (K) & (K) & $(M_{\odot}/\textrm{pc})$ & $(M_{\odot}/\textrm{pc})$ &
    (pc) & (pc) & (pc) & (pc) \\
(1) & (2) & (3) & (4) & (5) & (6) & (7) & (8) & (9) & (10) & (11) & (12) & (13) & (14) \\
\noalign{\smallskip}\hline
\noalign{\smallskip}
1 &  275.0401 & -14.5038  & 1.23 & $2.21\times 10^{21}$ & $0.21\times 10^{21}$ & 20.3 & 0.3 & 18.2 & 26.1 & 0.45 & 0.25 & 0.39 & 0.11 \\
2  &  275.0024 & -14.4626 & 1.07 & $3.24\times 10^{21}$ & $0.79\times 10^{21}$ & 19.8 & 0.3 & 25.2 & 27.0 & 0.31 & 0.03 & 0.32 & 0.03 \\
3  & 274.7068 & -14.1739  & 1.30 & $4.04\times 10^{21}$ & $1.03\times 10^{21}$ & 19.9 & 0.9 & 114 & 116 & 1.13 & 0.41 & 0.94 & 0.33 \\
4  & 275.1924 & -14.6405  & 1.25 & $6.33\times 10^{21}$ & $1.18\times 10^{21}$ & 19.6 & 0.4 & 46.6 & 41.4 & 0.35 & 0.08 & 0.31 & 0.09 \\
5 & 275.2809 & -14.6266   & 2.09 & $5.97\times 10^{21}$ & $1.18\times 10^{21}$ & 20.1 & 0.3 & 71.7 & 72.5 & 0.38 & 0.10 & 0.43 & 0.36 \\
$\cdots$ & $\cdots$ \\
\noalign{\smallskip}\hline
\end{tabular}

\medskip
\small
\begin{minipage}{\textwidth}
\setlength{\parindent}{0pt}
\setlength{\parskip}{2pt}
\textbf{Notes.} Column descriptions: (1) Filament identification number; (2), (3) Centroid equatorial coordinates (J2000) of the filament skeleton; (4) Filament (skeleton) length; (5) Median surface density along the crest; (6) Standard deviation of surface density along the crest; (7) Median dust temperature along the crest; (8) Standard deviation of dust temperature along the crest; (9), (10) Linear mass density measured from the left and right sides of the filament, respectively; (11) Median width measured from the left side; (12) Standard deviation of width (left side); (13) Median width measured from the right side; (14) Standard deviation of width (right side). All widths are measured as the full width at half maximum (FWHM) of the radial surface density profiles perpendicular to the filament skeleton.
\end{minipage}
\end{table*}

\label{lastpage}

\end{document}